\documentclass[a4paper,fleqn]{cas-dc}

\usepackage[numbers]{natbib}
\def\tsc#1{\csdef{#1}{\textsc{\lowercase{#1}}\xspace}}
\tsc{WGM}
\tsc{QE}
\begin{document}
\let\WriteBookmarks\relax
\def\floatpagepagefraction{1}
\def\textpagefraction{.001}

\shorttitle{Quantum dynamics via a hidden Liouville space}    

\shortauthors{G. O. Ariunbold}  

%\title [mode = title]{A Cascade Superradiance Model} 

\title [mode = title]{Quantum dynamics via a hidden Liouville space} 

%\tnotemark[] 


    %orcid=0000-0003-0430-7256

\author[1]{Gombojav O. Ariunbold}[orcid=0000-0003-0430-7256]

\cormark[1]

\fnmark[1]

\ead{ag2372@msstate.edu}

\ead[url]{www.ariunbold.physics.msstate.edu}

%\credit{asdds}

\affiliation[1]{organization={Department of Physics and Astronomy},
            addressline={Mississippi State University}, 
            city={Starkville},
%          citysep={}, % Uncomment if no comma needed between city and postcode
            postcode={39762}, 
            state={MS},
            country={USA}}

%\author[<aff no>]{<author name>}[]

% Footnote of the second author

% Email id of the second author
%\ead{}



%\affiliation[<aff no>]{organization={},
  %          addressline={}, 
    %        city={},
%          citysep={}, % Uncomment if no comma needed between city and postcode
      %      postcode={}, 
         %   state={},
           % country={}}

\cortext[1]{Corresponding author}



\begin{abstract}
Quantum dynamics for arbitrary system are traditionally realized by time evolutions of wave functions in Hilbert space and/or density operators in Liouville space. However, the traditional simulations may occasionally turn out to be challenging for the quantum dynamics, particularly those governed by the nonlinear Hamiltonians.
In this letter, we introduce a nonstandard iterative technique where
time interval is divided into a large number of discrete subintervals with an ultrashort duration;
and the Liouville space is briefly expanded with an additional (virtual) space only within these subintervals. 
We choose two-state spin raising and lowering operators for virtual space operators because of their simple algebra. This tremendously reduces the cost of time-consuming calculations.
We implement our technique for an example of a charged particle in both harmonic and anharmonic potentials. The temporal evolutions of the probability for the particle being in the ground state are obtained numerically and compared to the analytical solutions. 
We further discuss the physics insight of this technique based on a thought-experiment.
Successive processes intrinsically 'hitchhiking' via virtual space in discrete ultrashort time duration, are the hallmark of our simple iterative technique.
We believe that this novel technique has potential for solving numerous problems which often pose a challenge when using the traditional approach based on time-ordered exponentials.  
\end{abstract}


\begin{highlights}
\item A new iterative density-operator technique is introduced.
\item Quantum dynamics for the driven generalized quantum oscillators are studied both numerically and analytically.
\end{highlights}

% Keywords
\maketitle
%

% Main text

%%%%%%%%%%%%%%%%%%%%%%%%%%%%%%%%%%%%%%%%%%
%%%%%%%%%%%%%%%%%%%%%%%%%%%%%%
%%%%%%%%%%%%%%%%%%%%%
%%%%%%%%%%%%
%%%%%%
%%
%
%
%
%%%%%%%%%%%%%%%%%%%%%%%%%%%%%
%%%%%%%%%%%%%%%%%%%%%
%%%%%%%%%%%%
%%%%%%
%%
%

\section{Introduction}
Although, the standard approach based on time-ordered exponentials is extremely useful~\cite{Book, ScullyBook,Mukamel}, it may occasionally turn out to be challenging, particularly, in the case of revealing nonlinear quantum dynamics~\cite{PerinaBook,Wolf} that requires rigorous numerical simulations~\cite{Fisher,Braunstein87,Ari1999}. Quantum dynamics for arbitrary system are traditionally realized by time evolutions of wave functions in Hilbert space, which can also be expressed in terms of density operators in the Liouville space~\cite{ScullyBook,Mukamel}. 
In this letter, we introduce a new nonstandard iterative technique formulated as follows. 
(i) Finite time interval is divided into a large number of discrete subintervals with an ultrashort width. 
(ii) The Liouville space is expanded with an additional (i.e., virtual) space for this ultrashort time duration. The system's original Hamiltonian is, then, modified for the system's space plus virtual space, where the force terms are replaced with the virtual quantum operators. 
(iii) The density operator for the system is extracted by tracing over the virtual operator space. In principle, various virtual operators can be chosen depending on the specific quantum system. Here we choose two-state spin raising and lowering operators because of their simple algebra. 
In the next section, we present the standard approach using S-operator defined as time-ordered exponentials in Hilbert, and then, in the Liouville space. In section 3, we introduce our technique and implement it to the well-known example of a charged particle in a harmonic potential. The temporal evolutions of the probability for the particle being in the ground state are obtained by our technique and compared to the analytical solutions obtained using the standard S-operator. 
By extending this example, we perform numerical simulations for temporal evolutions for the ground state probability for the generalized systems governed by time-dependent nonlinear Hamiltonians. 
We further discuss the physics insight of this technique based on a thought-experiment, in which a large number of polarized atoms successively interact with a lossless cavity field. 
The last section is a conclusion.
%
%

%%%%%%%%%%%%%%%%%%%%%%%%%%%%%
%%%%%%%%%%%%%%%%%%%%%
%%%%%%%%%%%%
%%%%%%
%%
%
%
%
%%%%%%%%%%%%%%%%%%%%%%%%%%%%%
%%%%%%%%%%%%%%%%%%%%%
%%%%%%%%%%%%
%%%%%%
%%
%

\section{Standard approach}
In this section, the standard approach for quantum dynamics both in Hilbert space and the Liouville space is presented. 
We consider the system with the Hamiltonian given by
\begin{equation}
\hat{H}=\hat{H}_0+\hat{V}
\end{equation}
here $\hat{H}_0$ is the unperturbed (free) and $\hat{V}$ interaction Hamiltonians and we set $\hbar \equiv 1$.
\subsection{Quantum dynamics in Hilbert space}
We begin with the approach for the Hilbert space.
In the interaction representation, the rapid state evolution due to $\hat{H}_0$ is removed as
%
%\begin{equation}
%|\psi_{I}(t)\rangle={\rm e}^{i\hat{H}_0t}|\psi_S(t)\rangle
$|\psi_{I}(t)\rangle={\rm exp}{(i\hat{H}_0t)}|\psi_S(t)\rangle$,
%\end{equation}
%
where $|\psi_I(t)\rangle$ and $|\psi_S(t)\rangle$ are wave functions in the interaction and Schr${\rm \ddot{o}}$dinger representations, respectively.
%An arbitrary operator in these representations as $\hat{O}_I$ and $\hat{O}_S$ is given by
%
%\begin{equation}
%$\hat{O}_I(t)={\rm e}^{i\hat{H}_0t}\hat{O}_S{\rm e}^{-i\hat{H}_0t}$
%\end{equation}
%
%Therefore, 
Unitary transformation of initial state in the interaction picture is given as
%
%\begin{equation}
$|\psi_I(t) \rangle=\hat{U}(t)|\psi_I(0) \rangle$,
%\end{equation}
%
here unitary operator $\hat{U}(t)$ satisfies $\hat{U}(t)^\dagger\hat{U}(t)=\hat{1}$ and is expressed as
%
%\begin{equation}
%\hat{U}(t)={\rm e}^{i\hat{H}_0t}{\rm e}^{-i \hat{H}t}.
$\hat{U}(t)={\rm exp}{(i\hat{H}_0t}){\rm exp}{(-i \hat{H}t)}$.
%\end{equation}
%
Time evolution of $\hat{U}(t)$ can be derived from
%
%\begin{equation}
%i\frac{\partial\hat{U}}{\partial t}=\hat{V}_I(t)\hat{U}(t)
$i {\partial\hat{U}}/{\partial t}=\hat{V}_I(t)\hat{U}(t)$.
%\end{equation}
%
The Hamiltonian is in the interaction representation as $\hat{V}_I=\hat{V}_I^\dagger={\rm exp}{(i\hat{H}_0t)}\hat{V}{\rm exp}{(-i\hat{H}_0t)}$.
Choosing time interval between $t_2$ and $t_1$ $(t_1<t_2)$, unitary transformation is expressed as~\cite{Book}
\begin{equation}
|\psi_I(t_2)\rangle=\hat{S}(t_2,t_1)|\psi_I(t_1)\rangle
\end{equation}
with S-operator $\hat{S}(t_2,t_1)$. We divide time interval $t_2-t_1$ into $N$ sub-intervals with a width of $\Delta t$. At mid-time $\tau_j=t_1+(j-1/2)\Delta t$ in the $j$th interval, the S-operator is written as~\cite{Book}
\begin{equation}
\hat{S}\left( \tau_j+{\Delta t}/{2},\tau_j-{\Delta t}/{2} \right) = {\rm e}^{-i \hat{V}_I(\tau_j)\Delta t}
\end{equation}
where $N \rightarrow \infty $ and $\Delta t \rightarrow 0$ but $t_2 - t_1$ is finite.
Eq.(3) leads to the traditional time-ordered exponential given as~\cite{Book}
\begin{equation}
\hat{S}\left(t_2, t_1\right) = { \hat{T}} \left[ \prod_{j=1}^{N}{\rm e}^{-i \hat{V}_I(\tau_j)\Delta t} \right]= { \hat{T}} \left[{\rm e}^{-i \int_{t_1}^{t_2}\hat{V}_I(t)d t }\right]
\end{equation}
where time ordering for boson operators is defined as ${ \hat{T}} [\hat{O}_1(\tau_1)$ $\hat{O}_2(\tau_2) $ $\cdots$  $\hat{O}_N(t_N)]$ $=$ $\hat{O}_{p_1}(\tau_{p_1})$ $\hat{O}_{p_2}(\tau_{p_2})$ $ \cdots $ $\hat{O}_{p_N}(t_{p_N})$ with $\tau_{p_1} > \tau_{p_2} \cdots > \tau_{p_N}$.
%
%
%
%
%
%
%
%
%%%%%%%%%%%%%%%%%%%%%%%%%%%%%
%%%%%%%%%%%%%%%%%%%%%
%%%%%%%%%%%%
%%%%%%
%%
%
%
%
%%%%%%%%%%%%%%%%%%%%%%%%%%%%%
%%%%%%%%%%%%%%%%%%%%%
%%%%%%%%%%%%
%%%%%%
%%
%
%
\subsection{Examples}

As an example, we consider a driven harmonic oscillator. For that temporal evolutions of the ground state using S-operator are given in Eq.(4). Let a particle of a charge $q\equiv 1$, mass $m \equiv 1/2$  be in a harmonic potential $(\hbar \equiv 1)$. The driving electric field is $E(t) \equiv 1$, if $T> t >0$, and otherwise, it is zero and $\omega$ is the frequency of the oscillator. In the interaction picture, the Hamiltonian is written as
\begin{equation}
\hat{V}_I(t)=\hat{R}(t) \epsilon^\ast(t) + \hat{R}^\dagger(t) \epsilon(t) 
\end{equation}
where time-dependent operators are $\hat{R}(t)=\hat{b}(t)$ $=$ $\hat{b} {\rm e}^{-i \omega t}$ and $\hat{R}^\dagger(t)$ $=$ $\hat{b}^\dagger(t)$ $=$ $\hat{b}^\dagger {\rm e}^{i \omega t}$ and the force terms are
%
%\begin{equation}
%
$\epsilon(t)$ $=$ $\epsilon^\ast(t)$ $=$ $-\sqrt{{1}/{\omega}} E(t)$.
%
%\end{equation}
% 
The probability $p(T)$ for the particle to remain in the ground state $| \psi_0 \rangle $ after time $T$ is written as
\begin{equation}
p(T)=| \langle \psi_0 |\hat{S}(T,0) | \psi_0 \rangle|^2
\end{equation}
The probability amplitude is given by S-operator from Eq.(4) as
\begin{equation}
\langle \psi_0 |\hat{S}(T,0) | \psi_0 \rangle = \langle \psi_0 | { \hat{T}} {\rm e}^{-i\int_0^T \hat{V}_I(t') dt'} | \psi_0 \rangle 
={\rm e} ^{-iB(T)}
\end{equation}
where
%
%\begin{equation}
%
$B(T)=\int_0^T dtdt' \epsilon(t)^\ast G(t-t') \epsilon(t')$ 
%
%\end{equation} 
%
and $G(t)$ is the Green's function.  For this simple example, the Green's function is well known  
\begin{equation}
G(t ) =-i {\rm e} ^{-i\omega t} \theta(t). 
\end{equation}
Therefore, for the particle, its probability to remain in the ground state after time $T$ is analytically found to be as~\cite{Book}
\begin{equation}
p(T)=
{\rm exp} \left[ - \frac{4}{\omega^2}  {\rm sin}^2 \left( \frac{\omega T}{2} \right)\right] 
%{\rm exp} \left[ - \frac{4 t^2}{\Theta^2(t)}  {\rm sin}^2 \left( \frac{\Theta(t)}{2} \right)\right] 
%
\end{equation}
with pulse area $ \omega T$.

This example is the simplest case when a linear Hamiltonian is considered. That conveniently ensures to use the well known Green function in Eq.(8). 
However, in general, the Green's functions are mostly unknown and a laborious numerical method is often needed.
Next, we consider two more examples that use nonlinear Hamiltonians. 
%
%
%%%%%%%%%%%%%%%%%%%%%%%%%%%%%
%%%%%%%%%%%%%%%%%%%%%
%%%%%%%%%%%%
%%%%%%
%%
%
%
%\subsection{Examples: Driven nonlinear oscillators}
%
%%
%%%%%%%%%%%%%%%%%%%%%%%%%%%%%
%%%%%%%%%%%%%%%%%%%%%
%%%%%%%%%%%%
%%%%%%
%%
%
%
%
%
%
%
 The first example for the nonlinear Hamiltonian is a driven anharmonic oscillator. 
The Hamiltonian is given in the form in Eq.(5)~\cite{ScullyBook}
where degenerate two-boson nonlinear operators $\hat{R}(t)$ $=$ $\hat{b}^2(t)$ $=$ $\hat{b}^2 {\rm e}^{-2 i \omega t}$ and $\hat{R}^\dagger(t)$ $=$ $\hat{b}^{2}{}^\dagger(t)$  $=$ $\hat{b}^2{}^\dagger {\rm e}^{2 i \omega t}$ and the force terms $\epsilon(t)$ $=$ $\epsilon^\ast(t) $ $=$ $-\sqrt{{1}/{2\omega}} E(t)$ are assumed to have a similar form as in the harmonic oscillator case.
%
%
%
%%%%%%%%%%%%%%%%%%%%%%%%%%%%%
%%%%%%%%%%%%%%%%%%%%%
%%%%%%%%%%%%
%%%%%%
%%
%
%
%\subsection{Intensity-dependent oscillators}
%
%%
%%%%%%%%%%%%%%%%%%%%%%%%%%%%%
%%%%%%%%%%%%%%%%%%%%%
%%%%%%%%%%%%
%%%%%%
%%
%
%
%
%
%
%
 The second example for the nonlinear Hamiltonian is a driven intensity-dependent oscillator. 
The Hamiltonian is given in the form in Eq.(5)~\cite{Acta98}
with the intensity-dependent nonlinear boson operators  $\hat{R}(t)$ $ = $ $\hat{b}(t) \sqrt{\hat{b}^\dagger\hat{b}}$
and $\hat{R}^\dagger(t)$ $ = $ $\sqrt{\hat{b}^\dagger\hat{b}}\hat{b}^\dagger(t) $ and  $\epsilon(t)$ is assumed to be the same as before.
In section 3, we numerically solve for the probability time evolutions for these nonlinear systems and compare with the approximate analytical results. 
%
%
%
%
%
%
%
%
%
%%
%%%%%%%%%%%%%%%%%%%%%%%%%%%%%
%%%%%%%%%%%%%%%%%%%%%
%%%%%%%%%%%%
%%%%%%
%%
%
%
%%%%%%%%%%%%%%%%%%%%%%%%%%%%%
%%%%%%%%%%%%%%%%%%%%%
%%%%%%%%%%%%
%%%%%%
%%
%
%
\section{Nonstandard approach}
\subsection{Quantum dynamics in Liouville space}
Before introducing our technique, let us first replace the wave functions in Hilbert space with density operators in Liouville space~\cite{ScullyBook,Mukamel}. 
We recall that $\rho_{N}\equiv \rho(t_2) = |\psi_I(t_2) \rangle \langle \psi_I(t_2) |$ and $\rho_{0}\equiv \rho(t_1) = |\psi_I(t_1) \rangle \langle \psi_I(t_1) |$ from Eq.(2). 
Using S-operator in the $j$th interval from Eq.(3), we rewrite Eq.(2) in terms of density operators rather than wave functions as
\begin{equation}
\hat{\rho}_{j} = {\rm e}^{-i \hat{V}_I(\tau_j)\Delta t} \hat{\rho}_{j-1} {\rm e}^{i \hat{V}_I(\tau_j)\Delta t}
\end{equation}
here $\hat{\rho}_{j}=\hat{\rho}(\tau_j+{\Delta t}/{2})$ and $\hat{\rho}_{j-1}=\hat{\rho}(\tau_j-{\Delta t}/{2})$. 
In the traditional approach, to obtain $\hat{\rho}(t_2)$ at later time $t_2$ for any given initial state $\hat{\rho}(t_1)$ at $t_1$  Eq.(10) is repeatedly evaluated, where $\Delta t \ll 1$ and $N \gg 1$ but $t_2-t_1 = N \Delta t$  is finite. It is also important to note that Eq.(10) is the formal solution of the Liouville - von Neumann equation~\cite{ScullyBook}. 
\subsection{Quantum dynamics in Liouville space restructured with a virtual space}
From this point, we implement our new nonstandard approach, rather directly evaluating Eq.(10). 
As before, finite time interval $t_2-t_1$ is divided into $N$  discrete subintervals with an ultrashort width of $\Delta t$. The Liouville space is expanded with a two-state spin operator space for duration of $\Delta t$. The system's original Hamiltonian is, then, modified for the system's space plus spin space, where the force terms are replaced with the spin operators. The density operator for the system is extracted by tracing over the spin operator space.
In the $j$th interval with an infinitesimally short width of $\Delta t$, it is an acceptable ansatz where we replace the original Hamiltonian $\hat{V}_I(\tau_j)$ by new Hamiltonian expanding it with an additional virtual space $\hat{A}_j$ as
\begin{equation}
\hat{V}_{I}(\tau_j) \rightarrow \hat{V}_{I}(\tau_j) \otimes \hat{A}_j  
\end{equation}
where $\left[ \hat{V}_{I}(\tau_j), \hat{A}_j \right] =\hat{0}$.
For the sake of simplicity, $\hat{A}_j$ can be chosen to be a two-state spin operator defined as
\begin{eqnarray}
\hat{A}_j & = & | \alpha_j|^2 |\uparrow\rangle\langle \uparrow| + \alpha_j^\ast \beta_j  |\downarrow\rangle\langle\uparrow|
\nonumber\\
& + & \alpha_j \beta_j^\ast  |\uparrow\rangle\langle \downarrow|+| \beta_j |^2 |\downarrow\rangle\langle \downarrow|
\end{eqnarray}
with $| \alpha_j |^2 + | \beta_j|^2 =1$.
For this choice, the force terms are replaced with the raising $|\uparrow\rangle\langle \downarrow|$ and lowering $|\downarrow \rangle\langle \uparrow|$ operators specifically as
\begin{eqnarray}
&&\epsilon(\tau_j) {\rm e}^{i\omega \tau_j} \rightarrow \eta_j \alpha_j \beta_j^\ast |\downarrow\rangle\langle  \uparrow|  \nonumber\\
&&\epsilon^\ast (\tau_j){\rm e}^{-i\omega \tau_j} \rightarrow \eta_j^\ast \alpha_j^\ast \beta_j | \uparrow \rangle\langle  \downarrow|
\end{eqnarray}
 Thus, in the $j$th interval with $\Delta t$ width, this original Hamiltonian Eq.(5) can be replaced with a new Hamiltonian $\hat{V}_{A}(\tau_j)$, also known as the unified Jaynes-Cummings Hamiltonian~\cite{GBO1,GBO2} as
\begin{equation}
\hat{V}_A(\tau_j) = \eta_j^\ast \hat{R}(\tau_j) |\uparrow\rangle\langle  \downarrow|  +\eta_j\hat{R}^\dagger(\tau_j)|\downarrow\rangle\langle  \uparrow| .
\end{equation}
Instead of the original approach given by Eq.(10) for $\hat{\rho}_j$, we introduce an iterative relation for new density operator $\varrho_{j}$ using the modified Hamiltonian given in Eq.(14) as
\begin{equation}
\hat{\varrho}_{j} = {\rm Tr}_A \left[ {\rm e}^{-i \hat{V}_A(\tau_j)\Delta t} {\hat{A}_{j}} \otimes  \hat{\varrho}_{j-1}  {\rm e}^{i \hat{V}_A(\tau_j)\Delta t} \right] 
\end{equation}
Therefore, our goal is to demonstrate that the two density operators converge
\begin{equation}
\hat{\varrho}_N \simeq \hat{\rho}_N
\end{equation}
for the same pure initial state.  
%
%
%%%%%%%%%%%%%%%%%%%%%%%%%%%%%
%%%%%%%%%%%%%%%%%%%%%
%%%%%%%%%%%%
%%%%%%
%%
%
\subsection{A thought-experiment}
The essentials of our iterative technique are explained by the following thought-experiment. 
\begin{figure} [!ht]   
\begin{center}
\includegraphics[width=80mm]{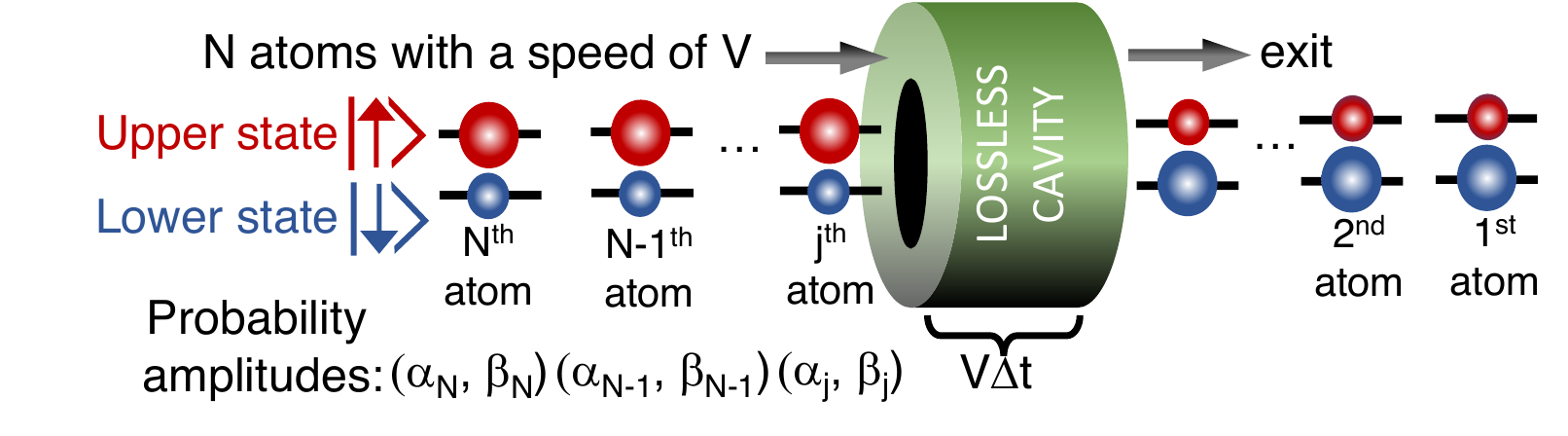}
\caption{A monokinetic beam of $N$ two-level atoms passing through a lossless cavity. Each atom interacts with the cavity field for a duration of $\Delta t$. The cavity field builds up to its final state $\hat{\varrho}_N$ (or $|\psi_I(t)\rangle$) from initial state $\hat{\varrho}_0$ (or $|\psi_I(0)\rangle$) after a finite time $t =N\Delta t$.  
}
\end{center}
\end{figure}
%%%%%%%%%%
As sketched in Fig.~1, let us consider a monokinetic beam consisting of individual two-level atoms. Each atom is prepared in arbitrary coherent superposition of the upper and lower states~\cite{Micromaser,Kien,Ari1999}. The atoms are then injected into a lossless cavity in a well controlled rate where only one atom at a time is present inside the cavity for duration $\Delta t$.  At the exit from the cavity the individual atoms are not intended to be measured. Total number of atoms is $N$ and the $j$th atom-field coupling constant is $\eta_j$. Although the present model can be generalized to multi-level atoms~\cite{Acta99,Acta98,Ari1999}, for the sake of simplicity, we consider only two-level atoms, where $|\uparrow \rangle$ and $|\downarrow \rangle$ are upper and lower atomic states, respectively. Correspondingly, $\alpha_j$ and $\beta_j$ are probability amplitudes for the $j$th atomic upper and lower states. Thus, as a result of numerically solving Eq.(15), the final cavity field state is evaluated from the existing initial quantum state in the cavity after time $t=\Delta t N$. 
For example, when atoms are prepared in the same phase then the cavity field evolves to the so-called superradiant state~\cite{Wolf,Ari1999,AriCSR,AriCs,Dicke,BookSR}. The mean number of photons created in the cavity (i.e., field intensity) is proportional to $N^2$ rather than $N$.
On the other hand, when each successive pair of atoms are prepared in perfectly out-of-phase, then the cavity field evolves to the sub-radiant state~\cite{Wolf,Ari1999}.
Moreover, we justify that time evolutions involve pure states after tracing over the virtual space operator. As demonstrated in our earlier work~\cite{Ari1999}, an initial coherent state given as $|\gamma_0\rangle$ evolves into $|\gamma (t) \rangle \simeq |\gamma_A + \gamma_0\rangle$, with $\gamma_A = -i\eta \alpha \beta^\ast  t$ at later time $t$. Therefore, the above statement that our technique maintains time evolutions for pure states is justified not only for infinitesimally short $\Delta t$ interval, but also for finite time $t$. 
%
%
%%%%%%%%%%%%%%%%%%%%%%%%%%%%%
%%%%%%%%%%%%%%%%%%%%%
%%%%%%%%%%%%
%%%%%%
%%
%
%
\subsection{Examples}
Here we employ our technique for the previous example for a charged particle in harmonic potential. We evaluate Eq.(15) using the Hamiltonian given in Eq.(14) both numerically and analytically, however, for the sake of simplicity, only for initial vacuum state. Thus, the modified Hamiltonian $\hat{V}_{A}(\tau_j)$ in Eq.(14) is written in terms of $\hat{R}(t) \equiv \hat{b}(t)$.
Comparing the Hamiltonian in Eq.(5) with the ansatz in Eq.(13), we obtain 
$\epsilon(\tau_j ) {\rm e}^{i\omega \tau_j} $ $ =$ $ - \eta_j \alpha_j \beta_j^\ast$ $=$ $- \eta_j\zeta_j^\ast$. The parameter $\zeta_j = \alpha_j^\ast \beta_j$ stands for a coherence between spin states. For  example, for parameters chosen to be as $\eta_j=1$, $|\alpha_j| = |\beta_j| =1/\sqrt{2}$, it is given by $- {\rm e}^{i \omega \tau_j}/2$ with $|\zeta_j| =1/2$.
 In the Fock state representation, the $j$th density matrix elements are $\rho_j(n,n')=\langle n | \hat{\rho}_j | n' \rangle $.
We numerically evaluate $\varrho_N(n,n')$ from Eq.(15), to obtain $\varrho_N(0,0)$, at $t_2-t_1=T$ with $t_1=0$ to compare the probability $p(T)$ given in Eq.(9). 
Eq.(9) is rewritten in terms of time-independent parameters $\eta_j=1$,  $|\zeta_j| =|\zeta|$ associated to the virtual operator space as
\begin{equation}
p(T) = {\rm exp}\left[ - \frac{4 |\zeta|^2}{\omega^2} {\rm sin}^2 \left(\frac{\omega T}{2}\right) \right]
\end{equation}
\begin{figure} [!ht]   
\begin{center}
\includegraphics[width=80mm]{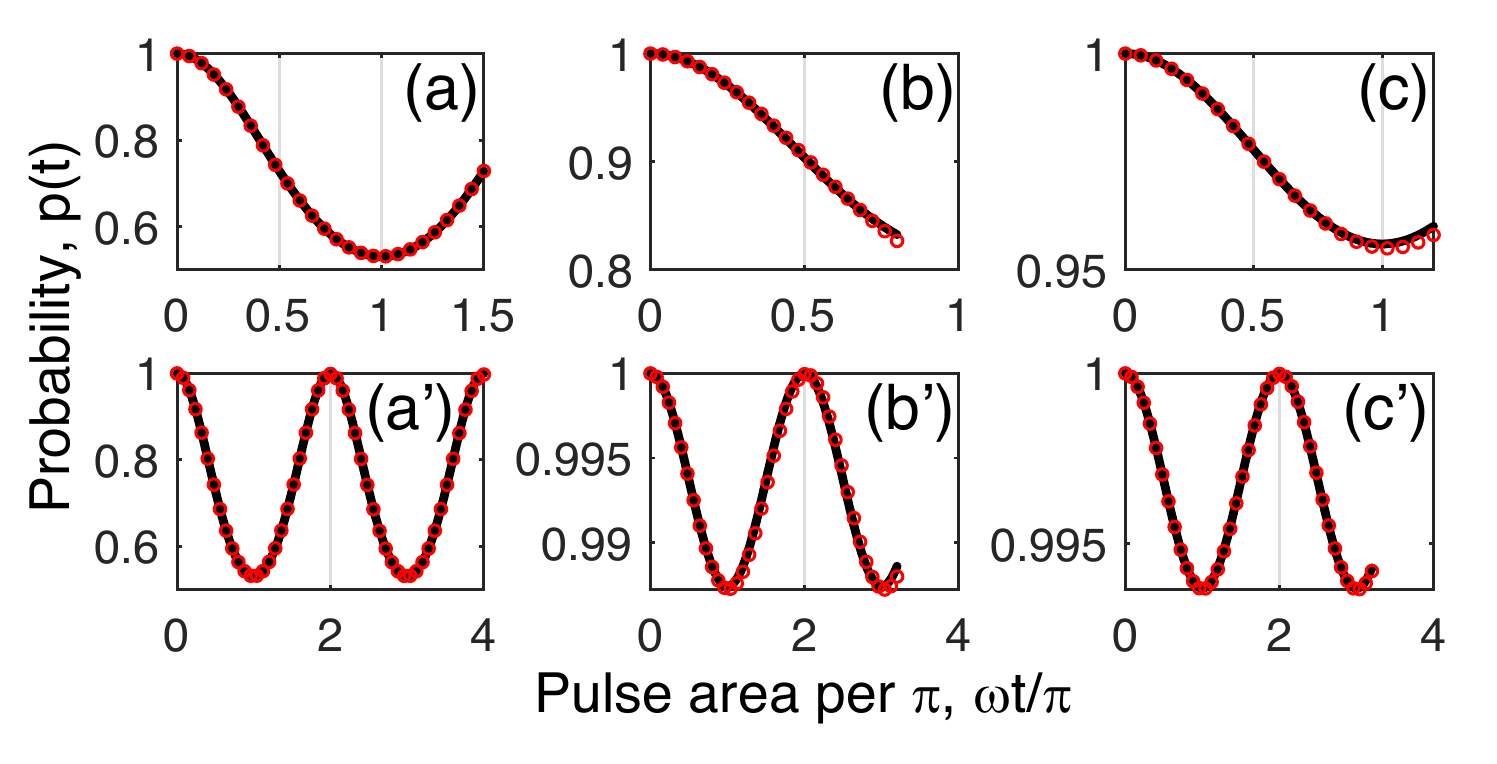}
\caption{Analytical (black curves) and numerical (red circles) results for temporal evolutions of the probability $p(t)$ for a charged particle driving by the external field being in the ground state as functions of $\omega t /\pi$. Left column: The system with a linear Hamiltonian. Middle column: The system with the degenerate two-boson Hamiltonian. Right column: The system with the intensity-dependent Hamiltonian.}
\end{center}
\end{figure}
%%%%%%%%%%
%%%%%%%%%%%%%%%%%%%%%%%%%%%%%%%%%%%%%%%%%%
%
%
In Fig.~2, the density matrix elements for $\varrho_N(0,0)$ (red circles) and $p(t)$ (black curves) are plotted as functions of pulse area per $\pi$, $\omega t /\pi$. 
For the plots in Figs.~2(a) and 3(a), the parameters include coherence $\zeta=1/2$, total number $N=3750$, width of the subintervals $\Delta t =0.001$, time $T=N \Delta t =3.75$ frequency $\omega = 2\pi/5$.
For the plot in Fig.~2(a'), except for the larger total number $N=10000$ and later time $T=10$, the rest of parameters remain the same as that given in Fg.~1(a).
%
%For Fig. 1(a'), the total number $N=10000$ width of the subintervals $\Delta t =0.001$, time $t=10$ frequency $\omega = 2\pi/5$.
%
\begin{figure} [!ht]   
\begin{center}
\includegraphics[width=80mm]{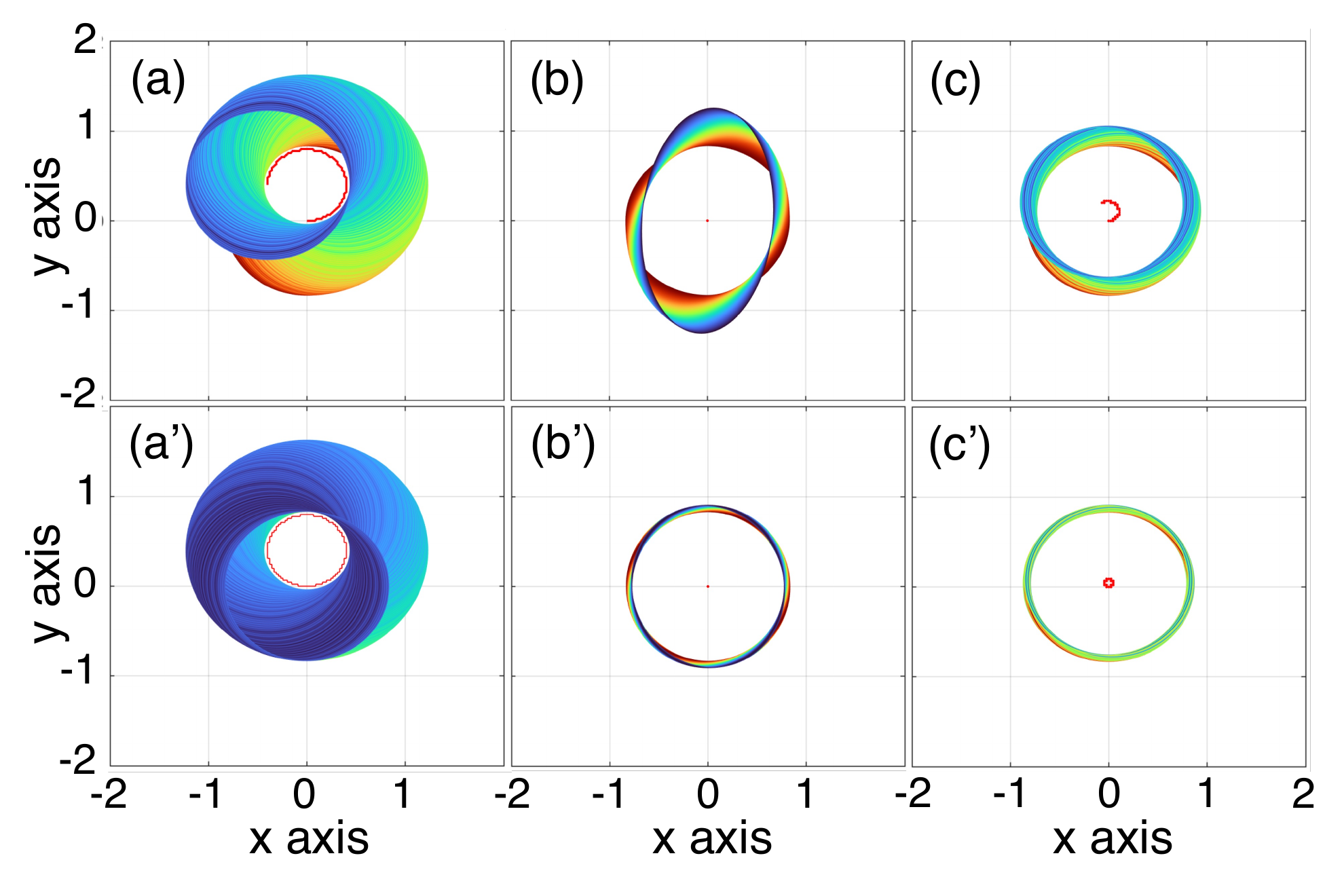}
\caption{Realizations of dynamics of quantum states from initial vacuum. Numerical calculations of  the temporal evolutions displayed by the contour plots of the Husimi Q-functions accompanied with the trajectories (red curves) of the centers of these contours. Left column: The system with a linear Hamiltonian. Middle column: The system with the degenerate two-boson Hamiltonian. Right column: The system with the intensity-dependent Hamiltonian. All parameters are the same as used in Fig.~2.
}
\end{center}
\end{figure}
%%%%%%%%%%
%%%%%%%%%%%%%%%%%%%%%%%%%%%%%%%%%%%%%%%%%%
%
%
In Fig.~3, the quasi-distributions given by the Husimi Q-functions are plotted. The Husimi Q-function~\cite{ScullyBook,PerinaBook,Wolf} is defined as $Q(x,y) = \langle \alpha | \hat{\varrho}_N   |\alpha \rangle /\pi$, here $\alpha = \sqrt{x^2 + y^2}$ ${\rm exp} [i \ {\rm atan}  (y/x) ]$. Because of coherent state representations, the Q-functions conveniently illustrate the coherent state as a displaced vacuum state with a perfect ring shape~\cite{Ari1999}, preserved for entire time. In Fig.~3, the red curves indicate the trajectories of displacements of the initial coherent state over time. These trajectories are the centers of single selected contour plots with the fixed value of the Q-functions at any given time $t$. For example, in Fig.~3(a), this trajectory follows a circle but is not yet complete circle opposite to that case in Fig.~3(a'). 
The parameters used for the plots in Fig.~3 (a,b,c,a',b',c') are the same as those used in Fig.~2(a,b,c,a',b',c'), respectively. 
In Fig.~2(b,b',c,c'), the realizations of quantum dynamics for nonlinear Hamiltonians with $\hat{R}(t)=\hat{b}^2(t)$ in (b,b') representing two-boson processes and $\hat{R}(t) = \hat{b}(t)\sqrt{\hat{b}^\dagger\hat{b}}$ in (c,c') representing intensity-dependent processes are demonstrated.
Similar to Fig.~2, the numerical results for $\varrho_N(0,0)$ are compared to approximate analytical expressions for time evolutions for the probabilities being in the ground state $p(T)$ after time $T$ in Fig.~3(b,c).
In the case of two-boson transition processes, the approximate analytical expressions are obtained to be 
\begin{equation}
p(T) \approx {\rm exp}\left[ - \frac{8 |\zeta|^2}{\omega^2} {\rm sin}^2 \left(\frac{\omega T}{2}\right) \right]
\end{equation}
For Figs.~2(b,c), the parameters are given as $N=8000$, $\Delta t =0.0001$, $T=0.8$, $ \omega=\pi $ and  $\omega T /\pi = 0.8$,
while for Figs.~2(b',c'), the parameters are the same as in (b) except for frequency $ \omega=4\pi $ and, thus, $\omega T /\pi = 3.2$.
In Fig.~3(b), the Q-functions display how the initial vacuum state with a ring shape is transformed to the significantly squeezed states with its signature oval shape~\cite{Yuen,Stoler,Ari1999,ScullyBook} for a slower process with a frequency of $\omega =\pi$. 
However, for the fast process with $\omega =4 \pi$, the state remains merely in vacuum state without observable squeezing. 
%If the state remains close to its ground state then the approximation Eq.(18) is strongly validated.   
%
%
%For Fig.~2(c), the parameters are given as $N=8000$, $\Delta t =0.0001$, $T=0.8$, $ \omega=\pi $ and $\omega T /\pi = 0.8$.
%
%
%For Fig.~2(c'), the parameters are given as $N=8000$, $\Delta t =0.0001$, $T=0.8$, $ \omega=4\pi $ and $\omega T /\pi = 3.2$.
%
Lastly, Figs.~2(c,c') and 3(c,c') represent the temporal evolutions for the Holstein-Primakoff ${\rm SU}(1,1)$ transformed states~\cite{HP,Acta98}. Similarly, the probabilities for slow (Fig.~2(c)) versus fast (Fig.~2(c')) processes are compared. It is important to note that the analytical formula for the probability for these processes is identical to Eq.(17). However, the deviation (i.e., displacement) is not as much pronounced as for coherent states (see, Fig.~3(c,c')). 
%Again, the statement that If the state remains close to its ground state then the approximation Eq.(18) is strongly validated, e.g., for fast processes.   
%
%In the next section, we conclude.
 %
\section{Conclusions}
In the standard approach, quantum dynamics for arbitrary system are realized by the time evolutions of wave functions in Hilbert space, which can also be expressed in terms of density operators in Liouville space. However, the standard quantum simulations may occasionally turn out to be challenging, particularly, for nonlinear dynamical systems.

In this letter, we introduce a new nonstandard iterative technique, formulated as follows. 
(i) A finite time interval is divided into a large number of discrete subintervals with an ultrashort width. 
(ii) The Liouville space is synthesized with an additional virtual space for ultrashort time duration and the quantum system's original Hamiltonian is modified accordingly. In particular, the force terms are replaced with virtual quantum operators. 
(iii) The density operator for the system is extracted by tracing over the virtual operator space. In principle, various virtual operators can be chosen depending on specific quantum system. For example, the simple algebra of using two-state spin raising and lowering operators reduces the cost of time-consuming calculations. 
After introducing our technique, we implement it to the well-known example of a charged particle in a harmonic potential. Temporal evolutions of the probability for the particle being in the ground state are obtained by the present technique and compared to the analytical solutions given by the standard approach. 
We further discuss the physics insight of this technique based on a thought-experiment.
Lastly, we perform numerical simulations for temporal evolutions for the ground state probability for generalized systems governed by the time-dependent nonlinear Hamiltonians. 
The quantum dynamics are realized by using the quasi-distributions. 

Successive processes implicitly 'hitchhiking' via virtual space for discrete ultrashort time duration, are the hallmark of our technique. We believe that this novel technique has potential for solving numerous problems 
 otherwise challenging to address using the standard approach based on time-ordered exponentials.
%

%\end{document}

%
%{\bf Acknowledgement}
%
%
%

%
%

\begin{thebibliography}{100}
\bibitem{Book}
P. Coleman, Introduction to many body physics, (Cambridge University Press, 2015)
%
\bibitem{ScullyBook}
M. Scully and S. Zubairy, Quantum optics, (Cambridge University Press, 1997)
%
%
\bibitem{Mukamel}
S. Mukamel, Principles of nonlinear optical spectroscopy, (Oxford University Press, 1995)
%
%
\bibitem{PerinaBook}
J. Perina, Quantum Statistics of Linear and Nonlinear
Optical Phenomena 2nd edn (Dordrecht: Kluwer 1991)
%
\bibitem{Wolf} 
% superradiand subradiant cite
L. Mandel, and E. Wolf, Optical Coherence and Quantum
Optics (Cambridge: Cambridge University Press, 1995)
%
\bibitem{Fisher}
% questionable to generalize
Fisher R A, Nieto M M and Sandberg V D 1984 Phys. Rev. D
{\bf 29} 110.
%
%
\bibitem{Braunstein87}
%
%pade approximants
S. L. Braunstein,  and R. I. McLachlan, 
Generalized squeezing
1987 Phys. Rev. A {\bf 35}
1659.
%
%
\bibitem{Ari1999}
G. Ariunbold, J. Perina, and Ts. Gantsog,
Nonclassical states in cavity with injected atoms
1999 J. Opt. B: Quantum Semiclass. Opt. {\bf 1} 219.
%
%
\bibitem{Acta98} 
G. Ariunbold, and J. Perina, 
Holsttein-Primakoff $SU(1,1)$ coherent state in micromaser under intensity dependent Janynes-Cummings interaction
1998 Acta Phys. Slov. {\bf 48} 315.
%
%
\bibitem{Acta99} 
G. Ariunbold, and J. Perina, 
Two-mode correlated states in cavity with injected atoms,
1999 Acta Phys. Slov. {\bf 49} 627.
%
%
\bibitem{GBO1}
%
% GBO Hamiltonian
D. Bonatsos, C. Daskaloyannis, and G. A. Lalassisis, 
Unification of Jaynes-Cummings models
1993,  Phys. Rev. A {\bf 47} 3448
%
\bibitem{GBO2}
%
% GBO space
P. Shanta, S. Chaturvedi, V. Srinivasan, and R. Jagannathan, 
Unified approach to the analogues of single-photon and multiphoton cohernet states for generalized bosonic oscillators
1994, J. Phys. A: Math. Gen. {\bf 27}  6433.
%
%
\bibitem{Micromaser}
%
D. Meschede, H. Walther and G. M${\rm \ddot{u}}$ller, 
One-atom maser
1985 Phys. Rev. Lett. {\bf 54} 551.
%
%
\bibitem{Kien} 
F. L. Kien, M. O. Scully, and H. Walther,  Generation of a coherent state of the micromaser field, 1993 Found. Phys. {\bf 23}
177.
%
%
\bibitem{Dicke} 
R. H. Dicke, 
Coherence in Spontaneous Radiation Processes,
Physical Review,
1954,  {\bf 93}, {99}
%
\bibitem{BookSR}
 {M. G. Benedict},
{Super-radiance: Multiatomic coherent emission}, 
 {(CRC Press, 1996)}
%
\bibitem{AriCSR}
%
G. O. Ariunbold, 
A cascade superradiance model,
2022, Phys. Lett. A, {\bf 452} 128468
%
%
\bibitem{AriCs} G. O. Ariunbold, V. A. Sautenkov, H. Li, R. K. Murawski, X. Wang, M. Zhi, T. Begzjav, A. V. Sokolov, M. O. Scully, and Yu. V. Rostovtsev,
Observations of Ultrafast Superfluorescent Beatings in a Cesium Vapor Excited by Femtosecond Laser Pulses, 2022, Phys. Lett. A, 428, 127945.
%
%
\bibitem{Yuen}
%two-photon coherent state or squeezed
H. P. Yuen, 
Two-photon cohernet states of the radiation field
1976 Phys. Rev. A {\bf 13} 2226
%
\bibitem{Stoler}
%two-photon coherent state or squeezed
%
D. Stoler, 
Equivalence Classes of minimum uncertainty packets,
1970 Phys. Rev. D {\bf 1} 3217
%
\bibitem{HP}
T. Holstein, and H. Primakoff, 
Field dependence of the intrinsic domain magnetization of a ferromagnet
1940, Phys. Rev. {\bf 58} 1098.
%
%
%\bibitem{Ari1} G. O. Ariunbold, M. M. Kash, V. A. Sautenkov, H. Li, Yu. V. Rostovtsev, G. R. Welch, and M. O Scully,
%Observation of Picosecond Superfluorescent Pulses in Rubidium Vapor Pumped by 100-Femtosecond Laser Pulses, Phys. Rev. A, 2010, {\bf 82}, 043421.
%
%\bibitem{AriOL} G. O. Ariunbold, V. A. Sautenkov, and M. O. Scully, Temporal coherent control of superfluorescent pulses,  Opt. Lett., 2012, {\bf 37}, 2400. 
%
%\bibitem{AriSok} G. O. Ariunbold, W. Yang, A. Sokolov, V. A. Sautenkov, and M. O. Scully, Superradiance in a Three-Photon Resonant Medium, Phys. Rev. A, 2012, {\bf 85}, 023424.
%
%\bibitem{AriFluct} G. O. Ariunbold, V. A. Sautenkov, and M. O. Scully, Quantum fluctuations of superfluorescence delay observed with ultrashort optical excitations, Phys. Lett. A, 2012, {\bf 376}, 335.
%\bibitem{Macfarlane}
%% q ananlog 
%A. J. Macfarlane, 
%On q-analoques of the quantum harmonic oscillator and quantum group $SU_q(2)$
%1989, J. Phys. A: Math. Gen. {\bf 22}, 4581.
%% q- analog
%\bibitem{Biedenharn}
%L. C. Biedenharn, 
%The quantum group $SU_q(2)$ and a q-analog of the boson operators,
%1989, J. Phys. A: Math. Gen. {\bf 22} L873.
%%
%\bibitem{AriNa} J. Thompson, C. W. Ballmann, H. Cai, Z. Yi, Yu. V. Rostovtsev, A. V. Sokolov, P. Hemmer, A. M. Zheltikov, G. O. Ariunbold, and M. O. Scully,
%Pulsed cooperative backward emissions from non-degenerate atomic transitions in sodium, New J. Phys. 2014, {\bf 16}, 103017.
%
%\bibitem{AriBack} G. O. Ariunbold, V. A. Sautenkov, and M. O. Scully, Ultrafast laser control of backward superfluorescence towards standoff sensing, Appl. Phys. Lett. 2014, {\bf 104}, 021114.
%
%
%
%
%
\end{thebibliography}
\end{document}